\documentstyle[12pt]{article}  
\begin{document}  \bibliographystyle{unsrt}

\vbox{\vspace{1.7in}}

\begin{center}

{\large \bf WAVELETS AND INFORMATION-PRESERVING TRANSFORMATIONS} \\[6mm]

Y. S. Kim\\
{\it Department of Physics, University of Maryland \\
College Park, Maryland 20742, U.S.A.}
\end{center}

\begin{abstract}
The underlying mathematics of the wavelet formalism is a representation
of the inhomogeneous Lorentz group or the affine group.  Within the
framework of wavelets, it is possible to define the ``window'' which
allows us to introduce a Lorentz-covariant cut-off procedure.  The window
plays the central role in tackling the problem of photon localization.
It is possible to make a transition from light waves to photons through
the window.  On the other hand, the windowed wave function loses
analyticity.  This loss of analyticity can be measured in terms of
entropy difference.  It is shown that this entropy difference can be
defined in a Lorentz-invariant manner within the framework of the wavelet
formalism.
\end{abstract}

\section{Introduction}
One of the still-unsolved problems in quantum mechanics is the
transition from classical light waves to photons.  Light waves are
classical objects, and their quantum counterparts are photons.  Then, are
photons light waves?  From the traditional theoretical point of view,
the answer is NO \cite{newwig49}.  However, this negative answer does
not prevent us from examining how close photons are to waves by
employing a new mathematical device called wavelet.  The word ``wavelet''
is relatively new in physics \cite{daube92,kaiser94},
but its concept was formulated in the 1960s \cite{aslak68}.  The
wavelet combines the traditional Fourier transformation with dilation
or ``squeeze'' and translational symmetries.  Since the squeezes and
translations are the basic symmetries in the Poincar\'e group, and since
the Fourier transformation is the standard language for the superposition
principle, the wavelet formulation of light waves gives a covariant
description of the superposition principle applicable to light waves.

Photons are relativistic particles requiring a covariant theoretical
description.  Wave functions in classical optics satisfy the superposition
principle and can be localized, but they are not covariant under Lorentz
transformations.  The wavelet formalism makes light waves covariant, and
this makes light waves closer to photons.  Furthermore, the formalism
allows us to make a quantitative analysis of the difference between these
two clearly defined physical concepts.  In this way, we can assert that
photons are waves with a proper qualification \cite{hkn96}.

In order to carry out this program, we need another important property
of wavelets: translation symmetry.  This symmetry allows us to introduce
the concept of ``window'' \cite{defaci90,szu92,kaiser93,hkn93}.  The
window allows us to keep a function defined within a specified interval
and let it vanish outside the interval or the window.  This finite
interval requires the concept of translation.  From the mathematical
point of view, this translational symmetry is very cumbersome and is
often misunderstood by physicists.  For instance, the translation does
not commute with the Lorentz boost.

However, it is possible to define the order of transformations to
preserve the information contained in the window \cite{hkn96}.  The
windowing process is a cut-off process, which leads to a loss of
information.  This information loss can be formulated in terms of the
entropy change.  It is shown that this entropy change can be formulated
in a covariant manner. In this report, we give a brief review of the
earlier work on this subject.  Sections \ref{waves}, \ref{windows},
\ref{photons} consist of review of the recent paper by Han, Kim, and
Noz \cite{hkn96}.  We report a new result in Sec. \ref{entropy}.  This
section deals with entropy.

\section{Light Waves and Wavelets}\label{waves}

For light waves, we start with the usual expression
\begin{equation}\label{2.1}
F(z,t) = {1\over \sqrt{2\pi }} \int g(k) e^{iku} dk \;,
\end{equation}
where $u = (z - t)$.  Even though light waves do not satisfy the
Schr\"odinger equation, the very concept of the superposition principle
was derived from the behavior of light waves.  Furthermore, it was
reconfirmed recently that light waves satisfy the superposition
principle \cite{aspect89}.  It is not difficult to carry out a spectral
analysis on Eq.(\ref{2.1}) and give a probability interpretation.

Before getting into the wavelet formalism, let us consider the expression
\begin{equation}\label{2.2}
A(z,t) = \int {1 \over \sqrt{2\pi\omega}} a(k) e^{iku} dk \;.
\end{equation}
This is the basic form we use in quantum electrodynamics, and is thus
very familiar to us.  We regard this as a classical quantity,
with an understanding that it will become the photon field after second
quantization.  This is a covariant expression in the sense that the norm
\begin{equation}\label{2.3}
\int \frac{|a(k)|^{2}}{2\pi\omega} dk \;.
\end{equation}
is invariant under Lorentz transformations, because the integral
measure $(1/\omega )dk$ is Lorentz-invariant.  It is possible to give a
particle interpretation to Eq.(\ref{2.2}) after second quantization.
However, $A(z,t)$ cannot be used for the localization
of photons.  On the other hand, it is possible to give a localized
probability interpretation to $F(z,t)$ of Eq.(\ref{2.1}), while it does
not accept the particle interpretation of quantum field theory.

Under the Lorentz boost:
\begin{equation}
z' = (\cosh\eta) z + (\sinh\eta) t \;, \qquad
t' = (\sinh\eta) z + (\cosh\eta) t \;,
\end{equation}
the variables $u$ and $k$ become $e^{-\eta}u$ and $e^{\eta}k$
respectively.  Thus, the Lorentz boost is a squeeze transformation in
the phase space of $u$ and $k$ \cite{kiwi87}.  The expression given in
Eq.(\ref{2.1}) is not covariant if $g(k)$ is a scalar function, because
the measure $dk$ is not invariant.  If $g(k)$ is not a scalar function,
what is its transformation property?  It was shown by Han, Kim, and
Noz \cite{hkn87} that we can solve this covariance problem by replacing
$F(u)$ and $g(k)$ by $F'(u)$ and $g'(k)$ respectively defined as
\begin{equation}\label{2.18}
F'(u) = \sqrt{ {p \over \sigma}} F(u) \;, \qquad
g'(k) = \sqrt{ {\sigma \over p}} g(k) \;,
\end{equation}
where $p$ is the average value of the momentum:
\begin{equation}
p = \frac{\int k|g(k)|^{2} dk}{\int |g(k)|^{2} dk} \;,
\end{equation}
which becomes $p = e^{\eta}$ under the Lorentz boost.
Then the functions of Eq.(\ref{2.18}) will satisfy Parseval's equation:
\begin{equation}\label{2.19}
\int |F'(u)|^{2} du = \int |g'(k)|^{2} dk
\end{equation}
in every Lorentz frame without the burden of carrying the multipliers as
given in Eq. (\ref{2.18}).  We can simplify the above cumbersome procedure
by introducing the form
\begin{equation}\label{2.20}
G(u) = {1 \over \sqrt{2\pi p}} \int g(k) e^{iku} dk \;.
\end{equation}
where the procedure for the Lorentz boost is to replace $p$ by $e^{\eta}p$,
and $k$ in $g(k)$ by $e^{-\eta}$.  As is shown in Ref. \cite{hkn87}, this
is a squeeze transformation.  This is precisely the wavelet form for
the localized light wave, and this definition is consistent with the form
given in earlier papers on wavelets \cite{daube92,aslak68}.

We are quite familiar with the expression of Eq.(\ref{2.1}) for wave
optics, and with that of Eq.(\ref{2.2}) for quantum electrodynamics.  The
above expression satisfies the same superposition principle as
Eq.(\ref{2.1}), and has the same covariance property as Eq.(\ref{2.2}).
It is quite similar to both Eq.(\ref{2.1}) and Eq.(\ref{2.2}), but they
are not the same.  The difference between $F(u)$ of Eq.(\ref{2.1}) and the
wavelet $G(u)$ is insignificant.  Other than the factor $\sqrt{\sigma}$
where $\sigma$ has the dimension of the energy, the wavelet $G(u)$ has the
same property as $F(u)$ in every Lorentz frame \cite{hkn87}.

However, there is a still a significant difference between $G(u)$ of
Eq.(\ref{2.20}) and $A(u)$ of Eq.(\ref{2.2}).  In Eq.(\ref{2.2}), the
factor ${1 /\sqrt{\omega}}$ is inside the integral and is a variable.
Thus the superposition principle is not applicable to $A(u)$ with $a(k)$
as a spectral function.  On the other hand, in Eq.(\ref{2.20}), the factor
${1 /\sqrt{p}}$ is constant.  Thus, the superposition principle is
applicable to $G(u)$ with $g(u)$ as a spectral function.

This difference disappears if the spectral functions $g(k)$ and $a(k)$ are
delta functions.   The difference becomes non-trivial as the widths of
these functions increase.  If the widths can be controlled, the difference
can also be controlled.  In so doing, it is essential that the spectral
functions vanish for $k = 0$ and for infinite values of $k$.  We can
achieve this goal by using the concept of window.

\section{Windows}\label{windows}
Here, the window means that the spectral function is non-zero within a
finite interval, and vanishes everywhere else.  There is a tendency to
regard this type of cut-off procedure as an approximation.  On the other
hand, we have to keep in mind that physics is an experimental science.
When we measure in laboratories, we do not measure functions, but we take
data points from which
functions are constructed.  It is well known that functions so constructed
can never be unique.  This is the limitation of accuracy in physics.

Thus, when we deal with a localized distribution, the window is a very
useful mathematical device.  Because of the inherent gap between a function
and a collection of data points, we do have the freedom of choosing a
windowed function for a localized distribution.  The problem here is the
question of covariance.  Let us choose a window in one Lorentz frame.
How would this window look in a different Lorentz frame?  Is this window
going to preserve all the information given in the initial frame.

In order to answer these questions, we have to examine the translation
symmetry of wavelets, particularly the translation combined with squeeze
transformations.  The problem here is that this affiance symmetry can
sometimes lead to information preserving windows and sometimes
non-preserving windows.  Let us look at this problem closely.

The problem is caused by the fact that squeeze transformations do not
commute with translations.  In the present case, the squeeze corresponds
to a multiplication of the variable $k$ by a real constant, while the
translation is achieved by an addition of a real number.  To a given
number, we can add another number, and we can also multiply it by another
real number.  This combined mathematical operation is called the affine
transformation \cite{gilmore74}.  Since the multiplication does not commute
with addition, affine transformations can only be achieved by matrices.
We can write the addition of $b$ to $x$ as
\begin{equation}\label{4.1}
\pmatrix{x' \cr 1} = \pmatrix{1 & b \cr 0 & 1} \pmatrix{x \cr 1} \;.
\end{equation}
This results in $x' = x + b$.  This is a translation.  We can represent
the multiplication of $x$ by $e^{\eta}$ as
\begin{equation}\label{4.3}
\pmatrix{x' \cr 1} = \pmatrix{e^{\eta} & 0 \cr 0 & 1} \pmatrix{x \cr 1} \;,
\end{equation}
which leads to $x' = e^{\eta}x$.  This is a squeeze transformation.

We are interested in combined transformations.  The translation
does not commute with the squeeze.  If the squeeze precedes the
translation, we shall call this the affine transformation of the first
kind, and the transformation takes the form
\begin{equation}\label{4.7}
x' = e^{\eta}x + b \;.
\end{equation}
If the translation is made first, we shall call this the affine
transformation of the second kind, and the transformation takes the form
\begin{equation}\label{second}
x' = e^{\eta}(x + b) \;.
\end{equation}
The distinction between the first and second kinds is not mathematically
precise, because the translation subgroup of the affine group is an
invariant subgroup.  We make this distinction purely for convenience.
Whether we choose the first kind or second kind depends on the physical
problem under consideration.  For a covariant description of light waves,
the affine transformation of the second kind is more appropriate, and the
inverse of Eq.(\ref{second}) is
\begin{equation}\label{4.11}
x = e^{-\eta}x'  - b = e^{-\eta} (x' - e^{\eta}b) \;.
\end{equation}
Therefore, the transformation of a function $f(x)$ corresponding to the
vector transformation of Eq.(\ref{second}) is
\begin{equation}\label{4.12}
f\left(e^{-\eta}x - b\right) = f\left(e^{-\eta}(x - e^{\eta}b) \right) \;.
\end{equation}

Next, does this translation lead to a problem in normalization as the
squeeze did?   The normalization integral does not depend on the
translation parameter $b$, but it does depend on the multiplication
parameter $\eta$.  Indeed,
\begin{equation}\label{4.14}
\int |f(e^{-\eta}x - b)|^{2} dx = e^{\eta} \int |f(x -b)|^{2} dx \;.
\end{equation}
In order to preserve the normalization under the affine transformation,
we can introduce the form \cite{daube92}
\begin{equation}\label{4.15}
e^{-\eta/2} f(e^{-\eta}x - b) \;.
\end{equation}
This is the wavelet form of the function $f(e^{-\eta}x - b)$.  This is of
course the wavelet form of the second kind.  The wavelet of the first kind
will be
\begin{equation}\label{4.16}
e^{-\eta/2} f\left(e^{-\eta}(x - b)\right) \;.
\end{equation}
Both the first and second kinds of wavelet forms are implicitly discussed
in the literature \cite{daube92}.  Here, we are using this concept in
constructing an information-preserving window under Lorentz boosts.

With this preparation, we can allow the function to be nonzero within the
interval
\begin{equation}\label{5.1}
a \leq x \leq a + w \;,
\end{equation}
while demanding that the function vanish everywhere else.  The parameter
w determines the size of the window.  The window can be translated or
expanded/contracted according to the operation of the affine group.  We
can now define the window of the first kind and the window of the second
kind.  Both windows can be translated according to the transformation
given in Eq.(\ref{5.1}).  The window of the first kind is not affected by
the scale transformation.  However, the size and location of the
window of the second kind becomes affected by the scale transformation
according to Eq.(\ref{second}).  We can choose either of these two windows
depending on our need.  The window of the first kind is useful when we
describe an observer with a fixed scope.

On the other hand, the window of the second kind is covariant and defines
the information-preserving boundary conditions \cite{hkn96}.
The width of this window is proportional to the average momentum, and the
ratio $w/p$ is a Lorentz-invariant quantity, where $w$ is the width of the
window.  Indeed, this information-preserving window will play an important
role in the photon localization problem.

\section{Photon Localization Problem}\label{photons}
Let us go back to Eq.(\ref{2.2}).  $A(u)$ is a classical amplitude, and
it becomes a photon field after second quantization.  If it is to be
localized, $a(k)$ must have a non-zero distribution, and $A(u)$ is
therefore a polychromatic \cite{glauber66,fearn89,campos90}.  $A(u)$ of
Eq.(\ref{2.2}) and $G(u)$ of Eq.(\ref{2.20}) are numerically equal if
\begin{equation}\label{ag}
a(k) = \sqrt{{k\over p}} g(k) \;,
\end{equation}
where the window is defined over a finite interval of $k$ which does not
include the point $k = 0$.  It is thus possible to jump from the wavelet
$G(u)$ to the photon field $A(u)$ using the above equation.  Furthermore,
since $k$ and $p$ have the same Lorentz-transformation property, this
relation is Lorentz-invariant.  Both $a(k)$ and $g(k)$ can be regarded as
distribution functions which can be constructed from experimental data.

However, the above equality does not say that $a(k)$ is equal to $g(k)$.
The intensity distribution of the localized light wave is not directly
translated into the photon-number distribution.  This is the quantitative
difference between wavelets and photons.  As we stated before, this
difference becomes insignificant when the window becomes narrow.  However,
as the window becomes narrower, the wavelet becomes more wide-spread.
The wavelet then becomes non-localizable.  This is why we are saying that
photons are not localizable.

However, there are no rules saying that $a(k)$ and $g(k)$ should be the
same.  As long as we can transform from one expression to other as in
Eq.(\ref{ag}, the transition from wavelets to photons can be carried out
within the window.  This point needs further investigation in the future.

\section{Entropy Formulation of the Information Loss}\label{entropy}
We now introduce the concept of entropy to deal with the information
loss due the windowing process.  We use in this report
the standard form for the entropy:
\begin{equation}
S = - \int \rho(k) \ln[\rho(k)] dk ,
\end{equation}
where $\rho(k)$ is the probability distribution function, with the
normalization condition
\begin{equation}
\int \rho(k) dk = 1 .
\end{equation}

If the Lorentz boost transforms $k$ into $e^{\eta}k$, the distribution
becomes widespread for positive values of $\eta$.  The normalization
integral becomes
\begin{equation}
\int e^{-\eta} \rho(e^{-\eta}k) dk = 1 .
\end{equation}
This normalization condition is form-invariant and is valid for all
normalizable probability distribution functions.  The Lorentz-boosted
entropy takes the form
\begin{equation}
S' = - \int e^{-\eta} \rho(e^{-\eta} k) \ln[e^{-\eta} \rho(e^{-\eta} k)] dk ,
\end{equation}
which becomes
\begin{equation}
S' = - \int \rho(k) \ln[\rho(k)] dk + \eta ,
\end{equation}
The effect of the Lorentz boost is very simple.  The boost simply
add the parameter $\eta$ to the original expression:
\begin{equation}\label{covent}
S' - S = \eta .
\end{equation}

The entropy difference between the analytic and windowed distribution
functions is
\begin{equation}\label{deltaS}
\Delta S = - \int \left\{\rho_{A}(k)\ln[\rho_{A}(k)]
- \rho_{W}(k)\ln[\rho_{W}(k)] \right\}  dk ,
\end{equation}
where $\rho_{A}(k)$ and $\rho_{W}(k)$ are the probability distributions
in the analytic and windowed forms respectively.  The integration of
this expression will produce a number.  The question then is whether
this is a Lorentz-invariant quantity.  Let us go back to Eq.(\ref{covent}).
The Lorentz-transformed entropy of the analytic form will produce
$\eta$, and so will the windowed form.  They will cancel each other.
Thus the expression for the entropy difference given in Eq.(\ref{deltaS})
is a Lorentz-invariant expression.

It is shown in Ref. \cite{hkn96} that an information-preserving window
can be defined.  In this report, we have shown that the information
loss due to the windowing process can also be defined in terms of a
the Lorentz invariance of the entropy difference.

\end{document}